\documentclass[aps,pra,twocolumn,reprint]{revtex4-1}

\usepackage{amsmath,hyperref,graphicx,color,ulem}

\hypersetup{
    colorlinks=true,       
    linkcolor=blue,          
    citecolor=blue,        
    filecolor=blue,      
    urlcolor=blue           
}

\newsavebox{\graphicsbox}

\usepackage{txfonts}

\def\um{\mbox{ $\mu$m}}
\def\nm{\mbox{ nm}}

\graphicspath{{./}{./figures/}}

\begin{document}

\title{Contact resistance and phase slips in mesoscopic superfluid atom transport} 

\author{S. Eckel}
\author{Jeffrey G. Lee}
\author{F. Jendrzejewski}
\author{C. J. Lobb}
\author{G. K. Campbell}
\author{W. T. Hill, III}
\email{wth@umd.edu}
\affiliation{Joint Quantum Institute, National Institute of Standards and Technology and University of Maryland, Gaithersburg, Maryland 20899, USA}

\date{\today}

\begin{abstract}
We have experimentally measured transport of superfluid, bosonic atoms in a mesoscopic system: a small channel connecting two large reservoirs.  Starting far from equilibrium  (superfluid in a single reservoir), we observe first resistive flow transitioning at a critical current into superflow, characterized by oscillations.   We reproduce this full evolution with a simple electronic circuit model.  We compare our fitted conductance to two different microscopic phenomenological models. We also show that the oscillations are consistent with LC oscillations as estimated by the kinetic inductance and effective capacitance in our system.  Our experiment provides an attractive platform to begin to probe the mesoscopic transport properties of a dilute, superfluid, Bose gas.
\end{abstract}


\maketitle

\section{Introduction}

Transport phenomena in mesoscopic systems are characterized by the importance of quantum phase coherence.  In these systems, the length scale associated with the device is typically smaller than or comparable to the inelastic mean free path.  This can lead to a wide variety of different effects, including quantum conductance~\cite{VanWees1988} and quantum persistent currents in normal metal rings~\cite{Bleszynski-Jayich2009,Bluhm2009}.  Cold atomic gases typically have mean free paths longer than the system size, and thus mesoscopic transport effects are crucial.  For example, in degenerate Fermi gases, quantum conductance has been observed~\cite{Krinner2015}.  Here, we measure mesoscopic transport in a Bose-condensed gas and observe superflow below a critical current and resistive flow, possibly associated with the creation of vortex pairs, above this critical current.  The dynamics of this system is well described using a simple circuit model which captures the essential physics.  In turn, the circuit parameters can be used to search for a microscopic explanation or to inform a more full modeling of system with mean-field theory.

Our system consists of two large condensates, or reservoirs, connected by a channel that is long compared to the healing length~\footnote{In traditional condensed matter systems like superfluid helium and superconductors, the healing length is typically referred to as the coherence length.} of the condensate but small compared to the mean free path (Figure~\ref{fig:setup}).  A similar experiment with thermal bosons observed a ballistic (Sharvin) resistance~\cite{Lee2013}.  Experiments in an analogous experiment~\cite{Brantut2012,Stadler2012,Brantut2013,Krinner2013,Krinner2015,Husmann2015,Krinner2015a} with fermions observed the superfluid transition~\cite{Stadler2012} and thermoelectric effects~\cite{Brantut2013}.  Our system, prototypical of many mesoscopcic transport devices, is of theoretical interest~\cite{Gallego-Marcos2014,Nietner2014} because it may help lead to new cooling mechanisms~\cite{Papoular2012,Papoular2014} and observation of the superfluid fountain effect~\cite{Karpiuk2012}.  In addition, if the channel is in the one-dimensional regime it can be described with a Hamiltonian similar to a Luttinger liquid~\cite{Simpson2014} and could violate the Wiedemann-Franz law~\cite{Filippone2014}.  Because of the long length of our channel, we expect to see different transport effects compared to using either tunnel barriers or short weak links~\cite{Sols1994,Piazza2010}. Experiments with such junctions have observed quantum effects like macroscopic quantum self-trapping~\cite{Albiez2005}, the ac and dc-Josephson effects~\cite{Levy2007}, and the transition from tunneling junctions to weak links~\cite{LeBlanc2011}.  Similar experiments with weak links and tunnel junctions in rings have shown resistive flow in superfluids~\cite{Ryu2013,Jendrzejewski2014}, persistent currents~\cite{Ramanathan2011}, and discrete phase slips~\cite{Wright2013,Eckel2014,Eckel2014b}.

In the experiments reported here, we observe that while the phase difference between the condensates governs the superfluid transport, there is also large dissipation.  This dissipation is related to the creation of excitations in one of the reservoirs, an effect thought to contribute to the resistance~\cite{Jendrzejewski2014} but not conclusively shown.  Because the creation of excitations appears to occur not within the channel but at the interface with the reservoir, the model allows us to consider our dissipation as a ``contact resistance''.  Contact resistance is a hallmark of mesoscopic transport: because the channel length is smaller than the inelastic mean free path, any dissipation must occur in the contacts.

\section{Brief Experimental Description}

\begin{figure}
	\includegraphics[width = \columnwidth]{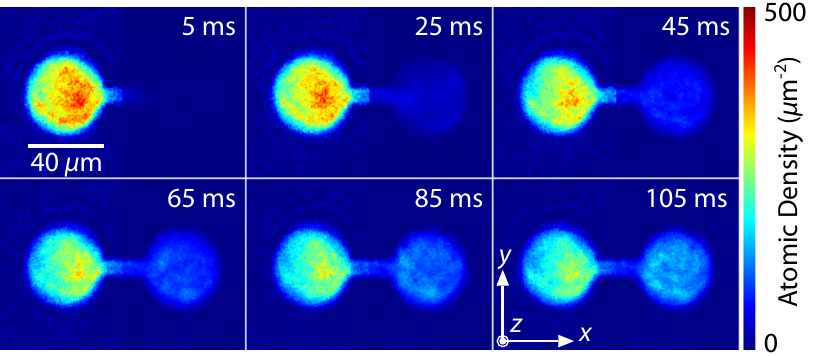}
	\caption{\label{fig:setup} {\it In-situ} images of a BEC of 495(16)~$\times 10^{3}$ $^{23}$Na atoms in the dumbbell-shaped potential for trapping parameters that yield an equilibrium 1-D density of atoms in the channel of 790(25)~$\mu$m$^{-1}$. The atoms are initially trapped in the left reservoir. At time $t=0$~ms, a gate is removed, and the atoms are allowed to flow freely between the reservoirs.}
\end{figure}

We start our experiments with all the atoms contained in one of the two reservoirs, as shown in Fig.~\ref{fig:setup}. The atoms are contained in this ``source'' reservoir by a ``gate'' potential.  When the gate potential is removed, the condensate starts to expand through the channel. The atoms that enter the channel are accelerated by the change in the interaction mean field energy and spray into the ``drain'' reservoir.   They then bounce off the walls of the reservoir and each other, causing them to thermalize.
	
\begin{figure}
	\includegraphics[width = \columnwidth]{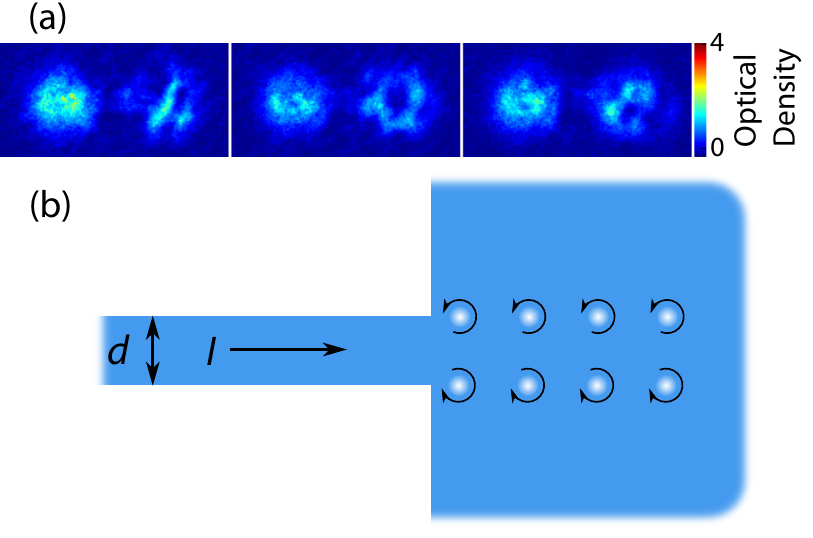}
	\caption{\label{fig:Composite_TOF} (a) Example 10~ms time-of-flight (TOF) images of the condensate after the atom number imbalance has reached equilibrium (612 ms after opening the gate). In almost all TOF images, we observe vortices, primarily in the initially empty reservoir, showing evidence of the Feynman mechanism for vortex production~\cite{Feynman1955}. (b) In the Feynman model, superfluid flows out of a channel into a reservoir. When the flow rate exceeds a critical value, vortex anti-vortex pairs are created. }
\end{figure}

Once a superfluid BEC is established in the drain reservoir, we expect that the supercurrent $I_s$ between the two reservoirs will be related to differences between their two phases.  If the rate is higher than the critical current of the channel, excitations will be created~\cite{Landau1941, Feynman1955, Raman1999, Crescimanno2000, Stiessberger2000, Fedichev2001, Recati2001, Astrakharchik2004, Chin2006, Arahata2009, Arahata2009a, Sykes2009, Wright2013a}.  Such excitations remove energy from the flow and eventually dissipate as heat.  Time-of-flight imaging shows vortices in the drain reservoir (Fig.~\ref{fig:Composite_TOF}a), and, while these are not the only type of excitation, these vortex excitations might play a key role.  The existence of vortices in this geometry is suggestive of the process described by Feynman~\cite{Feynman1955} and shown in Fig.~\ref{fig:Composite_TOF}b.  There, a channel of width $d$ carries atoms into a large reservoir, producing vortices at the corners where contact is made.  Refs.~\cite{Trela1967,Savard2011} successfully used the Feynman model to predict the critical velocity of superfluid liquid helium flowing through an orifice into a reservoir.

As the entire system moves toward equilibrium, the current drops below a critical value (the critical current) and dissipation decreases dramatically.  Any chemical potential imbalance that still exists at this time will result in number oscillations between the two reservoirs.  These oscillations are analogous to plasma oscillations in a superconducting junction~\cite{Orlando1991} and isothermal oscillations in superfluid liquid helium transport experiments~\cite{Robinson1951,Hartoog1981}.

Therefore, we expect to see two distinct behaviors, depending on the time after release.  We first expect to have a large amount of dissipation from the excitations that causes the mass imbalance to decay.  After the current drops below the critical value, we expect the resistive flow to decrease significantly and the current to oscillate.  Experiments with superfluid liquid helium in similar configurations produce qualitatively similar behavior (for examples, see Refs.~\cite{Allen1939,Hartoog1981}). 

\section{Experimental details}
\label{sec:exp_details}

Our experimental setup consists of a BEC of $\left|F=1,m_F=-1\right>$ $^{23}$Na atoms in an optical dipole trap (ODT).  Our BECs are created using standard laser cooling techniques, followed by evaporation in first magnetic then optical dipole traps.  The number of atoms in the condensate can be tuned between $10^{5}$ and $5\times10^{5}$ atoms, corresponding to equilibrium chemical potentials of $\mu_e/\hbar\approx 2\pi\times(500\mbox{ Hz})$ to $\mu_e/\hbar\approx 2\pi\times(1000\mbox{ Hz})$.  Vertical confinement is created using a red-detuned ($1064\nm$) ODT in the shape of a sheet. This sheet has a vertical trapping frequency of $\omega_z/2\pi \approx 529(2)$~Hz~\footnote{Unless stated otherwise, uncertainties are the uncorrelated combination of 1-$\sigma$ statistical and systematic uncertainties.  Stated uncertainties in atom number are the 1-$\sigma$ width of the atom number distribution, not the error in the mean.}.  The sheet also provides confinement in the plane of the dumbbell.  This residual confinement is characterized by a horizontal trapping frequency of $\approx 9$~Hz along the long axis of the dumbbell trap . 

In the horizontal plane, we use a direct intensity masking technique~\cite{Lee2015} to create the blue-detuned ($532\nm$) trap in the shape of dumbbell. This approach uses a Gaussian beam passing through a photomask in the shape of our desired potential.  The mask is imaged onto the atoms using the same optical system used to observe them.  The full-width, half max of the Gaussian beam is chosen to be approximately the end-to-end length of the dumbbell-shaped potential. 
In the plane, the reservoirs are nearly hard-walled with a diameter $D = 40(3)$~$\mu$m.  They are connected by a channel with length $l = 22(1)$~$\mu$m whose potential in $y$ is complicated by imperfections in the imaging process.  We empirically observe that the apparent Thomas-Fermi width of the channel $d=6.4(2)$~$\mu$m is independent of both total atom number $N$ and strength of the optical potential $U_m$.  However, the 1-D density of atoms in the channel, $n_\text{1D}$, does change with both $U_m$ and $N$.

There is also a variable-height gate potential that is used to block the channel in the middle.  The gate potential is created with a blue-detuned Gaussian beam that is scanned across the channel at 2~kHz using an acousto-optic deflector.  The combined height of this time-averaged gate potential and the static channel potential can be adjusted to be higher than the initial chemical potential of the source reservoir, blocking the channel. 

Due to the blue-detuned dumbbell trap having local minima outside of the region of interest, we adiabatically transfer the atoms into the source reservoir from an initial red-detuned ODT. The atoms are held in this configuration for at least 2~s to equilibriate, then the gate beam is turned off suddenly and the system is allowed to evolve.

At various times after opening the channel (typically ranging between 5~ms to $\leq1$~s), we count the number of atoms in each reservoir using partial-transfer absorption imaging (PTAI)~\cite{Ramanathan2012}.  To count the atoms in a given reservoir accurately, the transfer fraction is chosen to produce images with maximum optical densities between 1 and 2.  For $t$ near zero, images that count the number of atoms in the source reservoir will therefore have a small transfer fraction and cannot accurately count the small number of atoms in the drain.  To correct for this, a second image is taken with larger transfer fraction to accurately count the atoms in the drain.  Therefore, for each time observed, a pair of images is taken to determine the atom number in each reservoir.  In general, the resulting atom numbers from three or four pairs of images are averaged to determine the atom number imbalance at a given discharge time.

Because determining the 2D density from the optical density requires dividing by the transfer fraction, atomic densities determined from images with smaller transfer fractions will have larger noise.  This effect can be seen in Fig.~\ref{fig:setup}.  The images used to determine the atomic density in the source reservoir have a smaller transfer fraction and larger relative noise than the images used to determine the atomic density in the drain.

For these experiments, the temperature of our condensate is $\approx 100$~nK.  This temperature is determined by time-of-flight absorption imaging in a direction parallel to the plane of the dumbbell.  In this direction, the optical density of the thermal component is sufficiently large for detection.  In the plane of the dumbbell, the thermal component is too sparse to be effectively imaged.  We estimate the critical temperature of our condensate to be $\approx 500$~nK, and thus $>95$~\% of the atoms are in the condensed state.  Therefore, we expect little contribution to the bulk transport from the remaining 5~\% thermal atoms.  Moreover, the thermal cloud exists in regions outside the dumbbell (but still confined by the sheet potential), so we expect reasonable thermal contact between the two reservoirs.

The process of discharge represents a conversion of energy from chemical potential to both kinetic energy (in the form of collective excitations) and eventually to thermal energy.  We can estimate the maximum temperature increase of the system by considering the total energy of condensate $E \sim N\mu$.  The difference in the total energies of the initial state ($\mu/k_B \approx 30$ to 80~nK, depending on parameters) and final state ($\mu/k_B \approx 20$ to 50~nK) per particle represents the maximum increase in temperature.  For the parameters of our system, we expect this increase to be $\approx 10$ to 20~nK.

\section{Data and circuit model}

\begin{figure}
	\includegraphics{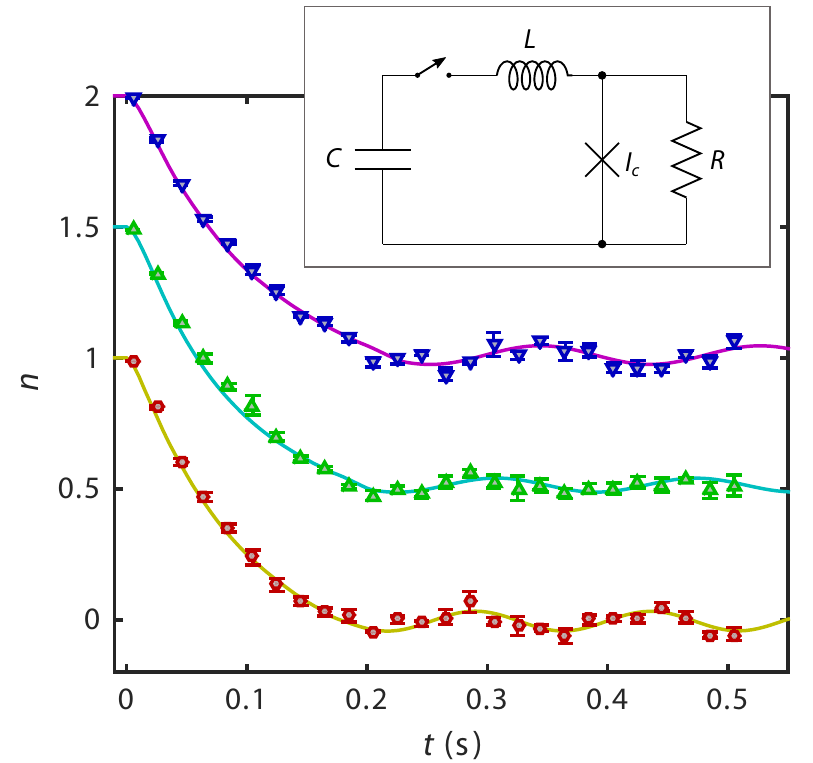}
	\caption{\label{fig:discharge} Normalized atom number imbalance between the two reservoirs vs.~time, for a total atom number of $472(22) \times 10^3$. The three plots shown are for different one-dimensional densities of atoms in the channel $n_\text{1D}$: red circles, 790(25)~$\mu$m$^{-1}$; green triangles, 665(16)~$\mu$m$^{-1}$; and blue, inverted triangles 599(17)~$\mu$m$^{-1}$. For clarity, decay measurements are artificially offset vertically by 0.5. The solid curves are fits to the expected dynamics from the circuit shown in the inset (see text).}
\end{figure}

Figure~\ref{fig:discharge} shows the atom number imbalance between the two reservoirs as a function of time after the gate potential is switched off. Here we define $N_e$, the equilibrium number of atoms in either reservoir.  We further define $\Delta N$ as the number imbalance, making $N_e+\Delta N$ ($N_e-\Delta N$) the number of atoms in the source (drain) reservoir. The plotted value, $n=\Delta N/2N_e$, is the normalized atom number imbalance, and can vary between $-1$ and 1 [where $n>0$ ($n<0$) represents more atoms in the source (drain) reservoir]. The evolution for three different $n_\text{1D}$ with $472(22)\times 10^3$ atoms are shown. As predicted, the atom number imbalance undergoes a short-time decay followed by an oscillation.  (No statistically significant conclusion can be drawn with respect to the timescale of the decay.)  We note that $n_\text{1D}$ (not shown) reaches an equilibrium quickly, typically $\lesssim20$~ms.


To fit this behavior, we model our system as a circuit that captures the essential physics described above.  Specifically, we consider the circuit in the inset of Fig.~\ref{fig:discharge}, which is a capacitor $C$~\cite{Lee2013} that discharges through an inductor $L$ connected in series to a resistance-shunted weak link (Josephson junction).  The capacitor represents energy stored in the chemical potential difference $\Delta\mu$ between the two reservoirs, while the inductor represents kinetic energy stored in the flow of atoms, both outside and inside the channel.  ($C$ is the only parameter we calculate {\it a priori}, see Sec.~\ref{sec:capacitance}.)  The weak link sets the critical current $I_c$ of the superfluid, and the resistor in parallel allows additional current to flow, but with dissipation.

There are three dynamical variables in this circuit: the number imbalance on the capacitor $\Delta N$ (defined above), the superfluid phase difference across the weak link $\gamma$, and the number current $I$.  The corresponding differential equations are
\begin{eqnarray}
	\frac{d(\Delta N)}{dt} & = & I \label{eq:curr_def} \\
	\hbar\frac{d\gamma}{dt} & = & V = R\left[I-I_c f(\gamma)\right] \label{eq:ac_eq} \\
	L\frac{dI}{dt} & = & -\left[\frac{\Delta N}{C} + V\right] = -\left[\frac{\Delta N}{C} + R\left(I-I_c f(\gamma)\right)\right]\ , \label{eq:kirchoff}
\end{eqnarray}
where $I_s = I_c f(\gamma)$ is the current-phase relationship of the weak link, $V$ is the voltage across the resistor and weak link, and $\hbar=h/2\pi$.  Equation~\ref{eq:curr_def} defines the current, Eq.~\ref{eq:ac_eq} is the AC Josephson law, and Eq.~\ref{eq:kirchoff} is Kirchhoff's law for voltage around the full circuit.  
These equations are integrated numerically with three independent parameters $\tau=RC$, $\omega^2=1/LC$, and $I_c$, which are determined by fitting to the data.  The curves shown in Fig.~\ref{fig:discharge} are the best fits for these data assuming $f(\gamma) = \sin\gamma$.  (As long as $f(\gamma)$ is $2\pi$-periodic, different current-phase relationships do not significantly alter the curves.)  By calculating $C$, we are able to extract the values of $R$ and $L$ from the fitted parameters in our model, $\tau$ and $\omega$.

\subsection{The capacitance}
\label{sec:capacitance}

\begin{figure}
	\center
	\includegraphics{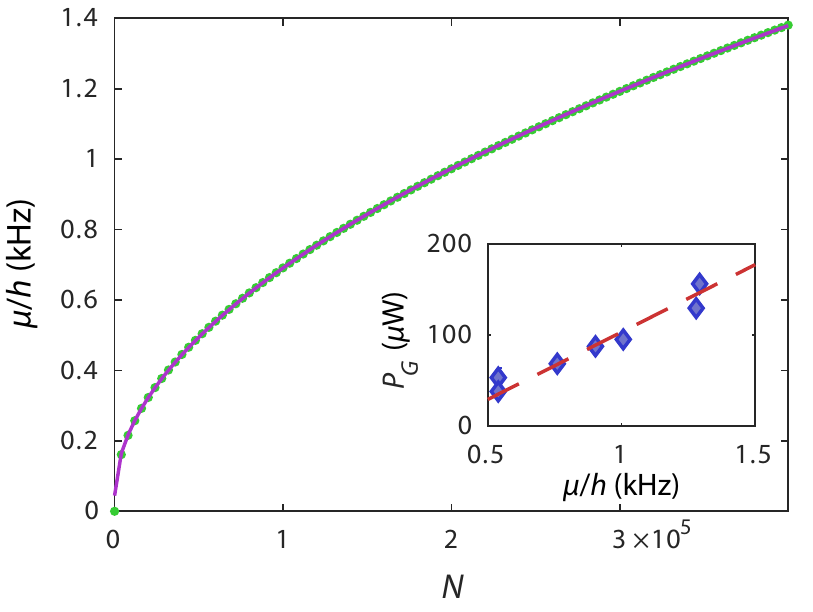}
	\caption{\label{fig:mu_vs_N}  Calculated reservoir chemical potential $\mu$ (green points) vs.~atom number $N$.  The magneta line shows the best fit power law.  Inset: measured power required in the gate beam to trap all the atoms in the source with Thomas-Fermi calculated chemical potential $\mu$.  The red, dashed line shows a linear fit and confirms the expected scaling.}
\end{figure}

We calculate the chemical capacitance following the methods used in Ref.~\cite{Lee2013} using our knowledge of the reservoir potential (see Appendix~\ref{sec:potential_modeling_reservoir}).  We calculate numerically, using the Thomas-Fermi approximation, how the chemical potential changes as a function of atom number for a fixed $U_m/\hbar = 2\pi\times1.8$~kHz.  The result is shown by the green points in Fig.~\ref{fig:mu_vs_N}.  The calculated $\mu$ is best fit by a power law of the form $\mu = \alpha + \beta N^\gamma$, where $\gamma=0.52(1)$,  $\alpha/h = 45(1)$~Hz, and $\beta/h = 1.60(2)$~Hz.   While $\gamma$ is roughly independent of $U_m$, the constants $\alpha$ and $\beta$ depend strongly on $U_m$.

By measuring the chemical potential relative to the height of our gate beam, we can experimentally verify this scaling behavior.  In particular, we measured the power of the gate beam that is required to prevent any atoms from spilling from the source reservoir into the drain for $U_m = 1.8(4)$~kHz.  The data is shown in the inset of Fig.~\ref{fig:mu_vs_N}, plotted versus the calculated Thomas-Fermi chemical potential.  The scaling should be linear, with an offset in $\mu$ that is roughly given by the offset in the channel potential (as the gate is placed across the channel and thus adds to it).  The fitted offset value of 300(100)~Hz agrees with our knowledge of the channel potential (see Appendix~\ref{sec:potential_modeling_channel}).

The chemical capacitance of the system can now be calculated from the difference in chemical potential between the two reservoirs:
\begin{eqnarray}
	\Delta\mu & = & \beta \left( \left( N_e + \Delta N \right) ^\gamma - \left( N_e - \Delta N \right)^\gamma \right) \\
	& \approx & \left(\beta N_e^\gamma\right) 2\gamma \frac{\Delta N}{N_e} = 2\gamma\frac{\mu_e-\alpha}{N_e}\Delta N
	\label{eq:chem_pot}
\end{eqnarray}
With $\gamma=0.52$, the linear approximation made here represents less than a 10\% error over a number imbalance up to approximately $\Delta N/N_e = \pm0.80$. With this approximation, the chemical capacitance of our system is then given by
\begin{equation}
	C = \frac{\Delta N}{\Delta\mu} \approx \frac{N_e}{2\gamma(\mu_e-\alpha)}.
	\label{eq:C}
\end{equation}

\subsection{The weak link}

The phase between the two condensates is approximated as being well defined in our model, as the variable $\gamma$ is the phase difference across the weak link.  The $2\pi$ periodicity of $f(\gamma)$ causes the supercurrent to oscillate when the chemical potential difference across the weak link $V$ is large.  This is the ac-Josephson effect.  Thus, the decay portion of the dynamics is similar to the self-trapped regime observed in Refs.~\cite{Albiez2005,Levy2007}.  However, due to the inductance, the high frequency (initially, about 500~Hz) oscillating current does not appear on the capacitor, and the observed total current does not contain {\it visible} Josephson oscillations, even in the model.

Given the nature of the discharge, one might wonder if approximating the superfluid flow as being through a weak link is accurate.  For example, at times before a superfluid is established in the drain reservoir, the phase difference should be undefined and the flow should be completely resistive.  The model works at these early times because the modeled superfluid flow is oscillating rapidly and averages to zero.  At later times when the oscillation is dominant, the phase profile in the drain is complicated by excitations, yet the model effectively averages the phase over the entire reservoir.  This two-mode approximation~\cite{Zapata1998} works because the local phase fluctuations in the drain due to excitations fluctuate at timescales that are smaller than the oscillation frequency.  A reasonable estimate of this timescale is the time for a phonon to traverse the reservoir, i.e., the diameter of the reservoir divided by the speed of sound.  For our system, this ranges from 10 to 20~ms, depending on the parameters.  Thus, over the period of oscillation, the fluctuations of the phase in the drain average out.

\subsection{The conductance}

\begin{figure}
	\includegraphics[width = \columnwidth]{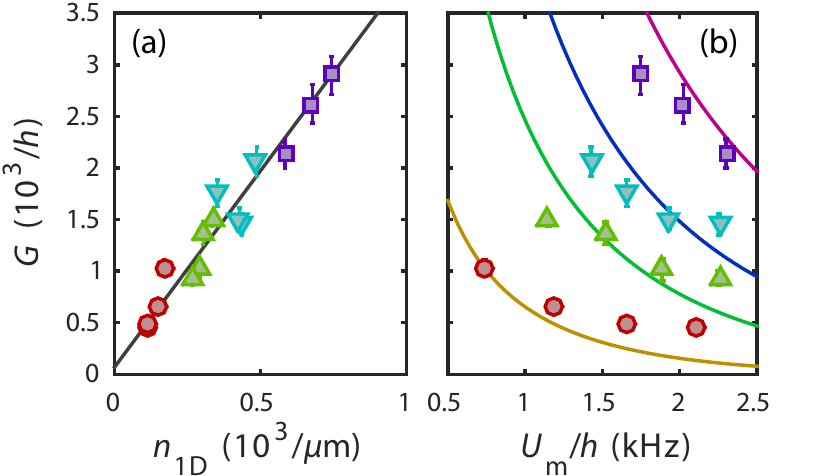}
	\caption{\label{fig:conductance_comparison} Measured conductance $G$ vs. (a) the one-dimensional density of atoms in the channel, $n_\text{1D}$, and (b) the height of the optical potential, $U_m$.  The solid black line is a linear fit, and the colored curves are the fit to the Feynman conductance (see text).  The experimental data are grouped by total atom number (violet squares $472(22) \times 10^3$; cyan, inverted triangle $331(11) \times 10^3$; green triangles $229(9) \times 10^3$; red circles $125(6) \times 10^3$).  The violet squares correspond to the data shown in Fig.~\ref{fig:discharge}.}
\end{figure}

Fig.~\ref{fig:conductance_comparison} shows our extracted conductances $G=1/R$.  In mesoscopic transport, the conductance is generally determined by the number of available single-particle transport modes~\cite{Datta1995}.  A fermionic system with a small number of transport modes (i.e., transverse single-particle modes that have an energy cutoff less than the chemical potential in either reservoir) displays quantum conductance, with each channel contributing a conductance of $h^{-1}$, where $h$ is Planck's constant.  Our particles are condensed bosons, and the absence of the Pauli exclusion principle implies no restriction on the conductance of an individual channel.  

We calculate the number of single particle transport modes that are available in the channel.  As described in Appendix~\ref{sec:potential_modeling_channel}, the exact form the channel potential is unknown, but is best described with $V\propto y^4$ with a non-negligible offset that has a component that is proportional to $U_m$ and a component that is constant.  Because the exact details of the channel remain unknown, we instead make a conservative estimate of the number of open transport channels by neglecting the offsets and considering the channel as being harmonic in the $\hat{z}$ direction and a square well with width $d$ in the $\hat{y}$ direction.  In this case, the energies of the single particle states with zero transverse momentum are given by
\begin{equation}
	E(n_y,n_z) = \frac{\pi^2\hbar^2n_y^2}{2md^2} + \left(n_z+\frac{1}{2}\right)\hbar\omega_z\ ,
\end{equation}
where $n_y$ and $n_z$ are the quantum numbers of the 2-D harmonic oscillator and $m$ is the mass of an atom.  The number of transport modes is then given by the number of combinations of $n_y$ and $n_z$ that satisfy
\begin{equation}
	E(n_y,n_z) < \mu\ .
\end{equation}
Our system thus has between 3 and 11 transport channels, depending on $\mu$ and $U_m$, yet we observe conductance of up to 2000 $h^{-1}$.

As shown in Fig.~\ref{fig:conductance_comparison}a, we observe that $G$ is directly proportional to $n_\text{1D}$.  Because $I=GV\approx G\Delta\mu$ during the decay, this dependence implies that the average velocity of the atoms in the channel is directly proportional to $\Delta\mu$.  Alternatively, because $G/h$ is unitless, $G/h = a n_\text{1D}$ implies the existence of a constant length scale $a=3.7(4)$~$\mu$m in the system.

Ref.~\cite{Jendrzejewski2014} suggested a simple model of phase-slip dominated conductance.  In this model, the conductance is attributed to the creation of excitations that carry $n_{1D}\xi$ atoms, where $\xi=\sqrt{\hbar/2m\mu}$ is the condensate healing length and thus is the relevant length scale for excitations.  The resulting conductance is $G_{PS}/h = 2n_\text{1D}\xi$.  We find a good fit with a single scaling parameter, $G = \alpha_{PS} G_{PS}$ where $\alpha_{PS} = 3.9(4)$, but only when using the local $\xi$ in the reservoirs (as opposed to the local $\xi$ in the channel).

\subsubsection{The Feynman model}
In the model that Feynman describes, superfluid flows at a constant rate through a channel into an infinite reservoir, as shown in Fig.~\ref{fig:Composite_TOF}b.  Above some critical velocity, the fluid can no longer sustain superflow, and vortices will be produced in pairs at either side of the channel, dissipating energy. Our experimental setup provides a unique opportunity to observe a system similar to that envisioned by Feynman. 

This model calculates the rate of vortex production $\gamma_p$ as
\begin{equation}
	\gamma_p = \frac{v^2 m}{2 \pi \hbar} = \frac{I^2 m}{2 \pi \hbar n_\text{1D}^2}
	\label{eq:vortex_rate}
\end{equation}
vortex pairs per second, where $v$ is the flow velocity.  In the reservoir, we can also estimate the energy of a pair of vortices separated by a distance $w$ as~\cite{Feynman1955}
\begin{equation}
	E_p = \frac{\pi n_\text{2D} \hbar^2}{m} \ln \left( \frac{w}{\xi} \right)\ ,
	\label{eq:vortex_energy}
\end{equation}
where $n_\text{2D}$ is the 2-D superfluid density. This calculation is for a vortex anti-vortex pair in a homogeneous BEC. To account for our finite geometry, we can use a method of images similar to that used in electro-magnetism~\cite{jackson1975classical} to numerically calculate the energy of a pair of vortices close to the wall of a circularly contained BEC. This calculation involves introducing image vortices outside of the reservoir such that the velocity field satisfies the boundary condition that it is tangent to the edges of the reservoir. Calculating the energy of the BEC with the resulting velocity field gives a correction factor of $\kappa \approx 1.7$ to the energy in Eq.~\ref{eq:vortex_energy}. Because the vortex pairs will be created at either side of the channel, we set $w=d$ in this calculation, where $d$ is the width of the channel.

From Eqs.~\ref{eq:vortex_rate} and~\ref{eq:vortex_energy}, the rate of energy dissipation in the system will be
\begin{equation}
	\label{eq:power}
	P = \kappa \gamma_p E_p = \kappa I^2 \frac{\hbar n_\text{2D}}{2 n_\text{1D}^2} \ln \left( \frac{d}{\xi} \right)\ .
\end{equation}
Equating this power with the power dissipated by a current through a resistor, $P = I^2 R_F$, we can now define the Feynman resistance in our system as
\begin{equation}
	R_F = \frac{1}{G_F} = \kappa \frac{\hbar n_\text{2D}}{2 n_\text{1D}^2} \ln \left( \frac{d}{\xi} \right).
\end{equation}
The critical current can be calculated by equating the power in the flow of atoms in the channel, $v \frac{1}{2}mn_\text{1D}v^2 = \frac{1}{2} mI^3 n_\text{1D}^{-2}$, to the power dissipated by vortices, Eq.~\ref{eq:power}~\cite{Feynman1955}.  For simplicity, we use the equilibrium value of the chemical potential $\mu_e$ to determine $\xi$.  For fitting to the experimental data, we use the experimentally determined densities $n_\text{1D}$ and $n_\text{2D}$.  Because $n_\text{1D} \approx d n_\text{2D}$, $G_F$ is approximately proportional to $n_\text{1D}$.

We fit this theory to the data through $G = \alpha_F G_F$, where $\alpha_F$ is a fit parameter.  The results of this fit are shown in Fig.~\ref{fig:conductance_comparison}b.  The model somewhat captures the trend, and the best fit parameter is $\alpha_F = 0.47(10)$.  The discrepancy from $\alpha_F=1$ may be due to the approximate nature of the Feynman model, the nature of our channel and/or the fact that there are other types of excitations not considered.

\subsection{The inductance and critical current}  

\begin{figure}
	\includegraphics[width = \linewidth]{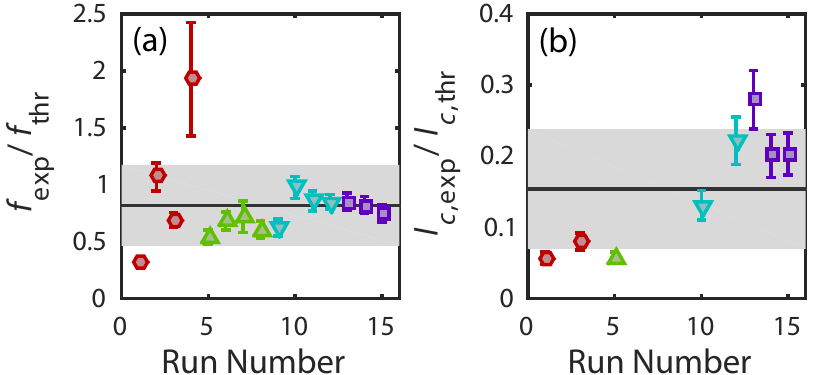}
	\caption{\label{fig:vc_T} A comparison between experimentally observed and theoretical values for both (a)~the period of LC oscillation and (b)~the critical velocity for vortex production in the system. For both plots, the data points are the values for each decay measurement (run), the solid line is a weighted mean of these values with a standard deviation represented by the shaded region. The experimental data are grouped by total atom number and delineated by color and symbol as in Fig.~\ref{fig:conductance_comparison}.}
\end{figure}

After the current drops below the critical current, the dissipation drops significantly, and we expect oscillations, described as plasma oscillations in Ref.~\cite{Papoular2014}.  Such oscillations represent energy coherently oscillating between kinetic energy (atoms moving through the channel) and chemical potential differences.  (Similar oscillations in liquid helium experiments are known as isothermal oscillations~\cite{Robinson1951} or plasma oscillations in Josephson junctions.)  Equivalently, it can be considered the first normal mode of our trap.  Such oscillations occur with statistical significance in roughly half of our decay measurements. All oscillations shown in Fig.~\ref{fig:discharge} are significant.

A good estimate of the kinetic inductance in our system is given by the kinetic inductance of the channel $L_c = m l/n_\text{1D}$~\cite{Lee2013}.  Using this and our calculation of the capacitance, we can estimate the expected oscillation frequency for our oscillator.  Figure~\ref{fig:vc_T}a shows a comparison of the experimentally measured frequency to that predicted by $1/\sqrt{L_c C}$.  We find the experimentally measured frequencies are, on average, $\approx15$\% lower than predicted.

Figure~\ref{fig:vc_T}b shows a comparison of the measured critical current, extracted from the amplitude of the oscillation, and the critical current predicted by the Feynman model.  We consider only decay measurements where the oscillation is statistically significant. (The amplitude of oscillation is a complicated function of the critical current, resistance, and capacitance.  Therefore, lack of a clear oscillation does not indicate a zero critical current.)  Our average measured critical current is approximately a factor of 5 below the predicted value.  This suggests that other excitations with lower critical velocities may be playing a role.

\section{Conclusion}

Our experiment provides an attractive platform to probe the mesoscopic transport properties of a dilute, superfluid, Bose gas through a small channel.  The mechanism of vortex production as described by Feynman~\cite{Feynman1955}, scaled by a factor of 2, predicts the general trend in our data (Fig.~\ref{fig:conductance_comparison}). Combined with the direct observation of vortices (Fig.~\ref{fig:Composite_TOF}), our experimental data suggests that the Feynman mechanism for vortex production plays a role in determining the conductance.  To conclusively show the relevance of the Feynman mechanism in similar mesoscopic cold-atom experiments, future experiments should use an initial condition where both reservoirs are partially filled and current bias the system by contracting one reservoir and expanding the second.  Lastly, when the current drops below the critical value, we observe plasma oscillations that are not visibly damped. 

Our experiment sets the stage for a number of other experiments.  First, with tighter channel potentials or fewer atoms, the number of open transport channels could be further reduced.  By controlling the initial imbalance, it might then become possible to see changes in the conductance as transport channels are opened or closed.  Moreover, this geometry could be used in studies of unique cooling mechanisms~\cite{Papoular2012,Papoular2014}, the superfluid fountain effect~\cite{Karpiuk2012}, or possibly seeing quantized conductance with bosons~\cite{Papoular2015}.  Because this experimental arrangement is capable of producing hundreds of vortex pairs, it could also prove useful in studying quantum turbulence~\cite{Paoletti2011} and perhaps the emergence of states like the Onsager vortex~\cite{Simula2014}.

\begin{acknowledgments}
The authors thank M. Edwards and T. Simula for useful discussions.  We thank W.D. Phillips for an extremely thorough reading of the manuscript.  This work was partially supported by ONR, the ARO atomtronics MURI, and the NSF through the PFC at the JQI.

S. Eckel and J.G. Lee contributed equally to this work.
\end{acknowledgments}

\appendix
\section{Modeling the dumbbell potential}
\label{sec:potential_modeling}
Here, we model the potentials of the reservoir and channel to determine simple functional forms and best-fit parameters.  We note, however, that in the present paper we do not rely on these best fit parameters in our understanding of the conductance or the Feynman model, as we have explicitly written our models to use the measured $n_\text{1D}$ of the channel, $n_\text{2D}$ and the apparent Thomas-Fermi width $d$ of the channel.  We include these models to better understand subtle effects (such as the effect of the residual transverse confinement on the capacitance of the system) and for possible future modeling of our experiment.

\subsection{The reservoirs}
\label{sec:potential_modeling_reservoir}
As described in Sec.~\ref{sec:exp_details}, the reservoir potential is made of three components: the harmonic confinement in the $\hat{z}$ direction, the residual transverse harmonic confinement in the $\hat{x}-\hat{y}$ plane, and the near square well potential in the $\hat{x}$-$\hat{y}$ plane created by the dumbbell potential itself.  The residual transverse harmonic confinement has its center in between the two reservoirs of the dumbbell.  As can be gleaned from Fig.~\ref{fig:setup}, this residual harmonic confinement is a perturbation on the overall potential, as it does not drastically alter the measured 2D density of atoms along the long axis of the channel.

As for the box portion of the potential, diffraction and imaging imperfections cause the hard-walled nature to be smoothed out.  If we were to approximate the aberrated point spread function of our imaging system as a Gaussian with $1/e^2$ radius $w$, then the resulting box portion of the reservoir potential will be given by the convolution of that point spread function with the optical mask used to generate the potential.  The resulting form of the reservoir potential is then
\begin{equation}
	V = \frac{1}{2}m\omega_x^2x^2 + \frac{1}{2}m\omega_y^2y^2 + \frac{1}{2}m\omega_z^2z^2 + \frac{U_m}{2}\left[1 + \text{erf}\left(\sqrt{2}\frac{r-r_0}{w}\right)\right]\ ,
\end{equation}
where $r=\sqrt{(x-y_c)^2+(x-y_c)^2}$ is the radial coordinate relative to the center of the reservoir, $x_c$ ($y_c$) is the $\hat{x}$ ($\hat{y}$) coordinate of the center of the reservoir, $r_0$ is the radius of the reservoir, erf is the error function, and $\omega_i$ is the trapping frequency in the $i$th direction.

We use this form of the potential along with the known number of atoms to calculate an expected 2D density.  We then fit, using all available equilibrium densities of both reservoirs, for the parameters $\omega_x$, $\omega_y$, $r_0$, and $w$.  The best fit values are $\omega_x/2\pi =9.1(9)$~Hz, $\omega_y/2\pi = 9.4(6)$~Hz, $w = 12(2)$~$\mu$m, and $r_0=27(2)$~$\mu$m.  Given the numerical aperture of the imaging stack, the expected values of $w$ is $\approx3$~$\mu$m.  The anomalously large value is most likely due to imaging aberrations, as described in detail in the next the subsection.

\subsection{The channel}
\label{sec:potential_modeling_channel}
If the imaging process were perfect with infinite resolution (i.e, no diffraction), we expect that the channel potential would be given by a square well with a width $d\approx 14\um$.  However, even in the absence of aberrations, our imaging system would produce an approximate square well with walls that changed from zero to the maximum height $U_m$ over a length scale $\approx 3\um$.  The potential is further complicated by the presence of optical aberrations, including both spherical aberration and astigmatism.  As a result, a region of sightly depleted density appears in the channel along the long axis nearly in the center.  This ``ridge'' is visible if one looks carefully at Fig.~\ref{fig:setup}.  These aberrations make effective modeling of the potential {\it a priori} virtually impossible, as it is unclear whether these aberrations are present in both imaging of the potential and subsequent imaging of the atoms, and, if so, in what relative quantities.

Instead, we choose to model the channel potential phenomenologically, using observables such as the {\it apparent} Thomas-Fermi width, integrated 1D density $n_\text{1D}$, and cross sectional profile (density vs. $\hat{y}$).  We find the potential is best described by $V \propto y^4$ potential, with the bottom of the potential having an offset given by $b U_m + V_0$, where $b = 0.15(2)$ and $V_0/h=223(30)$~Hz.  Here, $b$ represents the contribution to the offset due to imaging aberrations and $V_0$ represents a constant background potential, most likely due to a localized high point in the potential generated by the sheet beam.  Note that to accurately reproduce the data, one must take into account the 2D-3D crossover: if $\mu - V(y=0)\lesssim\hbar\omega_z$, we use only the ground state of the harmonic oscillator in the $\hat{z}$ direction; if $\mu - V(y=0)\gtrsim\hbar\omega_z$, we use the Thomas-Fermi solution in the $\hat{z}$ direction.

\bibliography{BECLibrary,TechLibrary,MesoscopicAndLiquidHeliumLibrary}

\end{document}